\documentclass[preprin,11pt]{elsarticle}
\usepackage[top=3cm,bottom=3.4cm,left=2cm,right=2cm]{geometry}

\makeatletter
\def\ps@pprintTitle{%
  \let\@oddhead\@empty
  \let\@evenhead\@empty
  \let\@oddfoot\@empty
  \let\@evenfoot\@oddfoot
}
\makeatother
\usepackage{xcolor}
\usepackage{bm}
\usepackage{hyperref}
\usepackage{amsmath}
\usepackage{amssymb}
\DeclareFontFamily{OT1}{pzc}{}
\DeclareFontShape{OT1}{pzc}{m}{it}{<-> s * [1.10] pzcmi7t}{}
\DeclareMathAlphabet{\mathpzc}{OT1}{pzc}{m}{it}
\newcommand{\nn}{\nonumber\\}

\numberwithin{equation}{section}
\begin{document}
\begin{frontmatter}

\title{Graviton vertex operators in the de Donder gauge
and BRST descent}

\author[socu]{Isao Kishimoto}
\ead{ikishimo@rs.socu.ac.jp}

\author[nwu]{Tomohiko Takahashi}
\ead{tomo@asuka.phys.nara-wu.ac.jp}

\author[nwu]{Mamiko Yamada}
\ead{yamada@asuka.phys.nara-wu.ac.jp}

\affiliation[socu]{
  organization={Center for Liberal Arts and Sciences,
  Sanyo-Onoda City University},
  addressline={Daigakudori 1-1-1},
  city={Sanyo-Onoda},
  state={Yamaguchi},
  postcode={756-0884},
  country={Japan}
}

\affiliation[nwu]{
  organization={Department of Physics, Nara Women's University},
  city={Nara},
  postcode={630-8506},
  country={Japan}
}

\begin{abstract}
We revisit the role of the trace sector of the graviton in covariant
string amplitudes by constructing the ghost-number-three graviton vertex
operator via BRST descent, imposing only the de Donder gauge condition
without requiring tracelessness. The resulting operator is well defined
even at zero momentum, while for nonzero momentum the
trace-dependent sector of the descent chain forms a trivial BRST descent
chain.
Evaluating
its disk one-point function in the presence of a D$p$-brane, we show
that the trace-dependent term is essential for reproducing the correct
graviton-brane coupling, canceling the Dirichlet-direction
contributions and leaving the expected coupling to the world-volume
components.
\end{abstract}

\begin{keyword}
BRST symmetry \sep graviton vertex operator \sep
descent equation \sep D-brane
\end{keyword}

\end{frontmatter}

%%%%%%%%%%%%%%%%%%%%%%%%%%%%%%%%%%%%%%%%%%%%%%%%%%%%%%
%\renewcommand{\thepage}{\arabic{page}}
%\renewcommand{\thefootnote}{\arabic{footnote}}
%\setcounter{page}{1}
%\setcounter{footnote}{0}
%\baselineskip=19pt plus 0.2pt minus 0.1pt
%%%%%%%%%%%%%%%%%%%%%%%%%%%%%%%%%%%%%%%%%%%%%%%%%%%%%%

%\tableofcontents

%%%%%%%%%%%%%%%%%%%%%%%%%%%%%%%%%%%%%%%%%%%%%%%%%%%%%%
\section{Introduction}

In covariant string theory, physical closed-string states are
characterized by the BRST cohomology~\cite{Kato:1982im}.
At the massless level, the
symmetric tensor state describes the graviton, and its polarization
tensor is conventionally taken to be transverse and
traceless~\cite{Polchinski:1998rq}.  The traceless condition is
appropriate for representing an external physical graviton, since
for nonzero momentum the trace component can be eliminated by a
residual gauge transformation.  Nevertheless, this does not
necessarily imply that the trace sector can be discarded in
covariant string amplitudes.

In an early analysis of one-loop vacuum amplitudes, Liu and
Polchinski pointed out, following an observation by Klebanov, that
the trace of the graviton in Feynman gauge contributes to the
factorization of the amplitude, even though it does not correspond
to an additional physical state~\cite{Liu:1987nz}.  This is
particularly suggestive at zero momentum, where the usual argument
for eliminating the trace by a residual gauge transformation ceases
to apply.  More generally, zero-momentum closed-string states are
known to exhibit subtleties in BRST cohomology that are absent at
generic momentum, especially in the graviton and dilaton
sectors~\cite{Bergman:1994qq,Belopolsky:1995vi}.

These observations motivate us to reconsider the graviton vertex
operator without imposing the traceless condition.  Instead, we
retain the trace of the polarization tensor and impose only the
de Donder gauge condition
\begin{align}
 k^\mu\varepsilon_{\mu\nu}
 -
 \frac{1}{2}k_\nu{\varepsilon^\mu}_\mu =0.
\label{eq:deDonder}
\end{align}
For nonzero momentum this description is gauge equivalent to the
usual transverse and traceless description, but it remains meaningful at
zero momentum and keeps track of the trace sector that is invisible
in the conventional representative of the physical graviton.

The BRST cohomology of closed strings contains states with
unconventional ghost numbers~\cite{Henneaux:1986kp}.  
In our previous work~\cite{Kishimoto:2024yuw}, we developed a
construction of closed-string vertex operators with unconventional
ghost numbers based on the Faddeev-Popov procedure.  The resulting
operators are naturally organized by the BRST descent equations
and, in particular, a
ghost-number-three vertex operator was constructed for the dilaton.
Its disk one-point function reproduces the dilaton tadpole with the
correct normalization, including the contribution associated with
additional terms~\cite{Kishimoto:2024yuw}.  The descent
construction has also recently been extended to superstring vertex
operators using integral forms on super Riemann
surfaces~\cite{Kishimoto:2026mkr}.

In this letter, we apply this framework to the graviton while
retaining the trace of the polarization tensor.  Starting from the
integrated graviton vertex in the de Donder gauge, we solve the BRST
descent equations without decomposing the polarization tensor into
traceless and trace parts.  We obtain a ghost-number-three vertex
operator that is well defined also at zero momentum.  For nonzero
momentum, we show that the trace-dependent
sector forms a trivial BRST descent chain,
consistently with the fact that the trace can
then be eliminated by a residual gauge transformation.  This
trivialization, however, cannot be extended directly to zero
momentum.

To make the role of the trace-dependent sector explicit, we evaluate
the disk one-point function in the presence of a D$p$-brane.
D-branes provide a particularly direct probe of the zero-momentum
coupling of closed-string fields~\cite{Polchinski:1998rq,
Polchinski:1995mt}.  We find that the conventional
matter part of the graviton vertex gives the difference between the
traces of the polarization tensor in the Neumann and Dirichlet
directions, whereas the trace-dependent term gives their sum.  The
Dirichlet components therefore cancel, leaving precisely the
expected coupling to the world-volume components of the graviton.
The resulting disk amplitude agrees with the standard D$p$-brane
coupling, including its overall normalization.  Thus, although the
trace-dependent sector is BRST trivial at nonzero momentum,
retaining it is essential for obtaining the correct zero-momentum
coupling in this representation.

\section{Graviton vertex operators in the de Donder gauge}

We start with the integrated vertex
operator for a graviton, which can be expressed as a two-form:
\begin{align}
 {\omega_2}^0
 = i dz\wedge d\bar z\, 
 \varepsilon_{\mu\nu}\partial X^\mu \bar{\partial}X^\nu e^{ik\cdot X}.
\end{align}
Here, $X^\mu(z,\bar{z})$ denotes the string coordinate field, and
$\varepsilon_{\mu\nu}$ is a symmetric
polarization tensor.
For the conventional graviton vertex operator, one usually imposes
the on-shell condition $k^2=0$ together with the transverse and
traceless conditions. In the present paper, however, we do not impose
the tracelessness condition and instead require the polarization tensor to satisfy
the de Donder gauge condition \eqref{eq:deDonder}, together with $k^2=0$.
This allows us to retain the
trace of the polarization tensor, whose treatment becomes nontrivial at zero
momentum.

For later use, we separate the polarization tensor into the
traceless part and the trace-dependent part.  For nonzero momentum $k^\mu$,
we introduce an auxiliary null vector $\bar{k}^\mu$ satisfying 
$\bar{k}\cdot k=1$ and $\bar{k}^2=0$. We then write
\begin{align}
&\varepsilon_{\mu\nu}=
\varepsilon'_{\mu\nu}+
\varDelta \varepsilon_{\mu\nu},
\label{eq:decomp_ep}
\end{align}
where $\varepsilon'_{\mu\nu}$ gives the usual transverse and traceless
graviton polarization, defined by
\begin{align}
&\varepsilon'_{\mu\nu}=
\varepsilon_{\mu\nu}
-\frac{1}{2}\left(
k_{\mu}\bar{k}_\nu+\bar{k}_\mu k_\nu\right) {\varepsilon^\lambda}_\lambda,
\end{align}
and $\varDelta \varepsilon_{\mu\nu}$ is given by
\begin{align}
 \varDelta \varepsilon_{\mu\nu}
=\frac{1}{2}\left(
k_{\mu}\bar{k}_\nu+\bar{k}_\mu k_\nu\right) {\varepsilon^\lambda}_\lambda,
\end{align}
and is therefore determined by the trace of the polarization tensor.
This decomposition is valid only for nonzero momentum,
since it relies on the existence of an auxiliary vector
$\bar k^\mu$.
For $k^\mu=0$, the
decomposition becomes ill-defined because $\bar{k}^\mu$ cannot be defined.

The vertex operator associated with
$\varDelta\varepsilon_{\mu\nu}$ is therefore
\begin{align}
 \varDelta{\omega_2}^0
=&
 i dz\wedge d\bar z\,
 \varDelta\varepsilon_{\mu\nu}
 \partial X^\mu\bar\partial X^\nu e^{ik\cdot X}
\nonumber\\
=&
 dz\wedge d\bar z\,
 \frac{i}{2}{\varepsilon^\lambda}_\lambda
 \left(
 k_{\mu}\bar{k}_\nu+\bar{k}_\mu k_\nu
 \right)
 \partial X^\mu\bar\partial X^\nu e^{ik\cdot X}.
\label{eq:Deltaw20}
\end{align}
We find that ${\varDelta\omega_2}^0$ is a total derivative:
\begin{align}
 \varDelta{\omega_2}^0
 =
 d  
 \left[
- dz\, \frac{1}{2}{\varepsilon^\lambda}_{\lambda}
\bar{k}_\mu \partial X^\mu e^{ik\cdot X}
 +
 d\bar z\, \frac{1}{2}{\varepsilon^\lambda}_{\lambda}
\bar{k}_\mu \bar\partial X^\mu e^{ik\cdot X}
 \right].
\end{align}
Therefore, for nonzero momentum, the trace-dependent part
of the integrated graviton vertex operator is a total
derivative and hence
does not contribute to amplitudes on closed worldsheets.
At zero momentum, however, the auxiliary vector
$\bar k^\mu$ cannot be defined, and the above argument
breaks down.
Consequently, the trace mode cannot be eliminated by this argument and must be
treated separately.

We now solve the descent equation without decomposing the
symmetric polarization tensor, keeping $\varepsilon_{\mu\nu}$ arbitrary
except for the de Donder condition \eqref{eq:deDonder}. 
We define
\begin{align}
\mathcal{G}^{\mu\nu}=\partial X^{\mu}\bar{\partial}X^{\nu} e^{ik\cdot X},
\quad
 A^\mu
 =
 \partial X^\mu e^{ik\cdot X},
\quad
 \tilde{A}^\mu
 =
 \bar\partial X^\mu e^{ik\cdot X}.
\end{align}
We denote the holomorphic and antiholomorphic reparametrization ghost
fields by $c(z)$ and $\tilde c(\bar z)$, respectively.
We then
obtain the following BRST transformation:
\begin{align}
{\bm\delta}_{\rm B}{\cal G}^{\mu\nu}=&
-\frac{i\alpha^{\prime}}{4}(\partial^2 c k^{\mu}\tilde{A}^{\nu}+\bar{\partial}^2\tilde{c} k^{\nu}A^{\mu})
\nonumber\\
&
+\Bigl(1+\frac{\alpha^{\prime}k^2}{4}\Bigr)(\partial c+\bar{\partial}\tilde{c})
{\cal G}^{\mu\nu}
+c\partial{\cal G}^{\mu\nu}+\tilde{c}\bar{\partial}
{\cal G}^{\mu\nu},
\label{eq:(204ik)}
%\\
%{\bm \delta}_{\rm B}(c\tilde{c}{\cal G}^{\mu\nu})
%&=-\frac{i\alpha^{\prime}}{4}c\tilde{c}(\partial^2 c k^{\mu}\tilde{A}^{\nu}
%+\bar{\partial}^2\tilde{c} k^{\nu}A^{\mu})
%+\frac{\alpha^{\prime}}{4}k^2
%(\partial c+\bar{\partial}\tilde{c})c\tilde{c}
%{\cal G}^{\mu\nu},
%\label{eq:(205ik)}
\end{align}
where the momentum $k^\mu$ is arbitrary.
Using the BRST transformation \eqref{eq:(204ik)} with $k^2=0$,
we find
\begin{align}
 \bm{\delta}_{\mathrm B} {\omega_2}^0
 =
 -d{\omega_1}^1 ,
\end{align}
where
\begin{align}
 {\omega_1}^1
 =&
i dz\left(\varepsilon_{\mu\nu}
 \tilde{c}\,\partial X^\mu\bar\partial X^\nu e^{ik\cdot X}
-\frac{\alpha'}{8}{\varepsilon^\lambda}_\lambda
\partial^2 c\,e^{ik\cdot X} \right)
\nonumber\\
&
-
i d\bar z\left(
\varepsilon_{\mu\nu}
 c\,\partial X^\mu\bar\partial X^\nu e^{ik\cdot X}
-\frac{\alpha'}{8}{\varepsilon^\lambda}_\lambda
\bar{\partial}^2 \tilde{c}\,e^{ik\cdot X}\right).
\label{eq:w11}
\end{align}
Furthermore, by computing the BRST transformation of \eqref{eq:w11},
we obtain
\begin{align}
 \bm{\delta}_\mathrm{B}{\omega_1}^1=d{\omega_0}^2,
\end{align}
where ${\omega_0}^2$ is given by
\begin{align}
 {\omega_0}^2=
ic\tilde{c}\varepsilon_{\mu\nu}\partial X^\mu \bar{\partial}X^\nu e^{ik\cdot X}
-i\frac{\alpha'}{8}{\varepsilon^\lambda}_\lambda (c\partial^2 c-\tilde{c}
\bar{\partial}^2 \tilde{c})e^{ik\cdot X}.
\label{eq:w02}
\end{align}
Finally, from \eqref{eq:(204ik)}, the zero-form ${\omega_0}^2$ is BRST invariant:
\begin{align}
 \bm{\delta}_\mathrm{B}{\omega_0}^2=0.
\end{align}
We have therefore obtained the complete 
BRST sequence for a graviton polarization tensor satisfying the de
Donder condition.

For nonzero momentum, in analogy with \eqref{eq:Deltaw20}
the trace-dependent parts of ${\omega_1}^1$ and ${\omega_0}^2$ are
obtained
by substituting the decomposition \eqref{eq:decomp_ep} into
\eqref{eq:w11} and \eqref{eq:w02}:
\begin{align}
 \varDelta {\omega_1}^1=&dz\,i{\varepsilon^\lambda}_\lambda\left\{
\tilde{c}\,\frac12 (k_\mu\bar{k}_\nu+k_\nu \bar{k}_\mu)
%\partial X^\mu \bar{\partial} X^\nu e^{ik\cdot X}
{\cal G}^{\mu\nu}
-\frac{\alpha'}{8} \partial^2 c e^{ik\cdot X}\right\}
\nn
&-d\bar{z}\,i{\varepsilon^\lambda}_\lambda
\left\{
c\,\frac12\,
(k_\mu\bar{k}_\nu+k_\nu\bar{k}_\mu)
%\partial X^\mu \bar{\partial} X^\nu e^{ik\cdot X}
{\cal G}^{\mu\nu}
-\frac{\alpha'}{8}\bar{\partial}^2 \tilde{c} e^{ik\cdot X}
\right\},
\\
\varDelta {\omega_0}^2
=&
\frac{i}{2}c\tilde{c}(k_\mu\bar{k}_\nu+k_\nu\bar{k}_\mu)
{\varepsilon^\lambda}_\lambda 
%\partial X^\mu \bar{\partial} X^\nue^{ik\cdot X}
{\cal G}^{\mu\nu}
-i\frac{\alpha'}{8}{\varepsilon^\lambda}_\lambda \left(
c\partial^2 c
-\tilde{c}\bar{\partial}^2 \tilde{c}\right)
e^{ik\cdot X}.
\end{align}
The BRST transformations of $A^\mu$ and
$\tilde{A}^\mu$ are given by
\begin{align}
 \bm{\delta}_\mathrm{B} A^\mu
=& -\frac{\alpha'}{4}\partial^2 c\,ik^\mu e^{ik\cdot X}
+\partial c
 \left(1+\frac{\alpha' k^2}{4}\right)A^\mu
+c\partial A^\mu
\nonumber\\
&
+\frac{\alpha'k^2}{4}\bar{\partial}\tilde{c}A^\mu
+\tilde{c}\bar{\partial}A^\mu,
\label{eq:dBA}
\\
 \bm{\delta}_\mathrm{B} \tilde{A}^\mu
=& -\frac{\alpha'}{4}\bar{\partial}^2 \tilde{c}\,ik^\mu e^{ik\cdot X}
+\bar{\partial} \tilde{c}
 \left(1+\frac{\alpha' k^2}{4}\right)\tilde{A}^\mu
+\tilde{c}\bar{\partial} \tilde{A}^\mu
\nonumber\\
&
+\frac{\alpha'k^2}{4}\partial c \tilde{A}^\mu
+c \partial \tilde{A}^\mu,
\label{eq:dBtildeA}
\end{align}
where the momentum is arbitrary,  as in
\eqref{eq:(204ik)}. % and \eqref{eq:(205ik)}.
Using \eqref{eq:dBA} and \eqref{eq:dBtildeA} with $k^2=0$,
we find that $\varDelta {\omega_1}^1$ can be expressed as
the sum of a BRST exact term and a total derivative,
while $\varDelta {\omega_0}^2$ is BRST exact:
\begin{align}
 \varDelta {\omega_1}^1=\,&
\bm{\delta}_\mathrm{B}
\left\{\frac{{\varepsilon^\lambda}_\lambda}{2}
\left(dz\,\bar{k}\cdot A-d\bar{z}\,\bar{k}\cdot \tilde{A}\right)
\right\}
\nonumber\\
&
\qquad
+d\left\{
-\frac{{\varepsilon^\lambda}_\lambda}{2}
\left(c\bar{k}\cdot A-\tilde{c} \bar{k}\cdot \tilde{A}\right)\right\},
\\
\varDelta {\omega_0}^2=\,&
\bm{\delta}_\mathrm{B}
\left\{
-\frac{{\varepsilon^\lambda}_\lambda}{2}
\left(c\bar{k}\cdot A-\tilde{c} \bar{k}\cdot \tilde{A}\right)\right\}.
\end{align}
Thus, for nonzero momentum, the trace-dependent contributions to the
descent sequence are trivial up to BRST-exact and total derivative
terms.

\section{Graviton vertex operators with ghost number three}

We next construct the graviton vertex operator with ghost number
three and its ascendants.  In our previous work~\cite{Kishimoto:2024yuw},
we constructed closed-string vertex operators with various ghost
numbers by reconsidering the gauge fixing of the conformal Killing
group $PSL(2,\mathbb{R})$ on the disk. In particular, fixing the
position of a single closed-string vertex leads, through the
Faddeev-Popov procedure, to the additional ghost insertion
\begin{align}
\frac{i}{4\pi}
\left(\partial c-\bar{\partial}\tilde c\right).
\label{eq}
\end{align}
For a  conventional matter vertex,
multiplying the unintegrated closed-string vertex by
this factor gives a vertex operator with ghost number three.
The normalization of this ghost insertion is fixed by the Faddeev-Popov
procedure rather than chosen by convention.

For the dilaton, however, we found that this simple construction
requires a modification. Since the ghost-number-two dilaton vertex
contains the ghost-dilaton contribution, multiplication by
\eqref{eq} alone does not yield a BRST-invariant
operator. An additional term is required to construct the
BRST-invariant dilaton vertex operator with ghost number three.

Applying  the same construction to the graviton, we obtain
the BRST-invariant graviton vertex operator with ghost number three
as follows:
\begin{align}
 {\omega_0}^3 =&
-\frac{1}{4\pi}(\partial c-\bar{\partial}\tilde{c})
c\tilde{c} \varepsilon_{\mu\nu}
\partial X^\mu \bar{\partial}X^\nu e^{ik\cdot X}
\nn
&
\hspace{-.7cm}+\frac{1}{2}\,{\varepsilon^\lambda}_\lambda\left\{
\frac{\alpha'}{16\pi}(\partial c-\bar{\partial}\tilde{c})
(c\partial^2 c-\tilde{c}\bar{\partial}^2 \tilde{c})e^{ik\cdot X}
+\varDelta {\omega_0}^3(h)
\right\}.
\label{eq:W03}
\end{align}
The last term in \eqref{eq:W03} is required to ensure BRST
invariance and is constructed, as in the dilaton case, using
operators that explicitly involve the string coordinate $X^\mu$.
In~\cite{Kishimoto:2024yuw},
the corresponding term
$\varDelta {\omega_0}^3$ for the dilaton
was constructed in this manner.
To make explicit the dependence on the function $h(x)$,
we denote the additional term for the graviton by
$\varDelta {\omega_0}^3(h)$.
Explicitly, it is given by
\begin{align}
\varDelta {\omega_0}^3(h)=\,&
\tilde{c}\bar{\partial}^2 \tilde{c} c C(h)
+c\partial^2 c \tilde{c}\tilde{C}(h)
\nonumber\\
&
-\frac{\alpha'}{2}(\partial c+\bar{\partial}\tilde{c})
(c\partial^2 c+\tilde{c}\bar{\partial}^2
\tilde{c})e(h).
\label{eq:dW03h}
\end{align}
Here the operators  $C(h)$, $\tilde{C}(h)$, and $e(h)$ are defined by
\begin{align}
&
 C(h)\equiv h(k\cdot X) X\cdot \partial X e^{ik\cdot X},
\quad
 \tilde{C}(h)\equiv h(k\cdot X)X\cdot \bar{\partial}X e^{ik\cdot X},
\nonumber\\
&
 e(h)\equiv h(k\cdot X) e^{ik\cdot X}.
\label{eq:Ce_def}
\end{align}
The function $h(x)$, introduced in the dilaton construction
of~\cite{Kishimoto:2024yuw}, is defined by
\begin{align}
 h(x)=\frac{1}{4\pi}\sum_{n=0}^\infty \frac{(d-3)!}{(d-2+n)!}(-ix)^n,
\label{eq:hxdef}
\end{align}
which satisfies the differential equation
\begin{align}
 (d-2) h(x)+ix h(x)+x h'(x)=\frac{1}{4\pi}.
\label{eq:bibuneq}
\end{align}
For convenience, 
we keep the spacetime dimension $d$ arbitrary throughout 
the intermediate calculation and set $d=26$ at the end.
The BRST variation can be evaluated in the same manner
as for the dilaton vertex operator in~\cite{Kishimoto:2024yuw}.
Using \eqref{eq:bibuneq}, all terms in the variation cancel, and
we obtain
\begin{align}
 \bm{\delta}_{\mathrm B}\,{\omega_0}^3=0.
\end{align}

Although the expression for $\varDelta{\omega_0}^3(h)$ is
somewhat complicated, it has the important advantage of remaining
well defined even at zero momentum.  For nonzero momentum
($k^\mu\neq0$), however, a much simpler expression can be obtained
by introducing an auxiliary null vector $\bar{k}^\mu$ satisfying
$k\cdot\bar{k}=1$ and $\bar{k}^2=0$.

The auxiliary-vector representation also makes it possible to show
explicitly that the trace-dependent part of the ghost-number-three
vertex operator is BRST exact. Using the decomposition
\eqref{eq:decomp_ep}, we write
\begin{align}
 {\omega_0}^3
 =
 \omega_{0{\rm T}}^{3}
 +\frac{1}{2}{\varepsilon^\lambda}_\lambda
 \omega_{0{\rm S}}^3 ,
\end{align}
where $\omega_{0{\rm T}}^3$ is the transverse-traceless part and
$\omega_{0{\rm S}}^3$ denotes the trace-dependent part.

To establish the BRST-exact form compactly, we introduce a new function
$F(x)$, defined in terms of $h(x)$ by
\begin{align}
F(x)
=&
\frac{1}{1-i\frac{d}{dx}}h(x)
=
\sum_{n=0}^{\infty}
i^n\frac{d^n}{dx^n}h(x)
\nonumber\\
=&
\frac{1}{4\pi}\sum_{n=0}^{\infty}
\frac{(d-4)!}{(d-3+n)!}
(-ix)^n.
\label{eq:Fxdef}
\end{align}
The inverse operator in this expression is understood formally,
and the last equality follows by rearranging the resulting double
series.
Applying the formal inverse operator $1/(1-id/dx)$ 
to \eqref{eq:bibuneq},
we obtain
\begin{align}
(d-3)F(x)+ix\left(1-i\frac{d}{dx}\right)F(x)
=
\frac{1}{4\pi}.
\label{eq:226ikv14}
\end{align}
The corresponding operator $\Phi_0^2(F,\bar{k})$ is defined by
\begin{align}
\Phi_0^2(F,\bar{k})=&
-\frac{4}{\alpha^{\prime}}c\tilde{c}\bar{\varepsilon}^{(-)}_{\mu\nu}(F)
\mathcal{G}^{\mu\nu}
-2i(\partial c+\bar{\partial}\tilde{c})
\nonumber\\
&\times
\left\{
c\bar{k}_{\mu}
F(k\cdot X)A^{\mu}+\tilde{c}\bar{k}_{\mu}F(k\cdot X)\tilde{A}^{\mu}
\right\},
\end{align}
where
the antisymmetric operator
$\bar{\varepsilon}_{\mu\nu}^{(-)}(F)(\cdots )$ is defined by
\begin{align}
\bar{\varepsilon}_{\mu\nu}^{(-)}(F)(\cdots )
=&\,
\bar{k}_\mu
F\left(-i\hat{k}_\lambda\frac{\partial}{\partial{k}_\lambda}\right)
\frac{\partial}{\partial{k}_\nu}(\cdots)\bigg|_{\hat{k}=k}
\nonumber\\
&
-\bar{k}_\nu
F\left(-i\hat{k}_\lambda\frac{\partial}{\partial{k}_\lambda}\right)
\frac{\partial}{\partial{k}_\mu}(\cdots)\bigg|_{\hat{k}=k}.
\end{align}
In this definition, the derivatives with respect to
$k^\mu$ are evaluated off shell, and
the on-shell condition $k^2=0$ is imposed only after the differentiation.
The auxiliary momentum $\hat{k}^\mu$ is treated as
independent of $k^\mu$ during the differentiation and is set equal
to $k^\mu$ afterward.
Using the BRST
transformation rules  given above, one can verify
directly that
\begin{align}
 \omega_{0{\rm S}}^3
 =
 \bm{\delta}_{\rm B}
 \left[
 \Phi_0^2(F,\bar{k})
 +\frac{i}{4\pi}(\partial c-\bar{\partial}\tilde c)
 \left(
 c\,\bar{k}\cdot A
 -\tilde c\,\bar{k}\cdot\tilde{A}
 \right)
 \right].
\label{eq:traceBRSTexact}
\end{align}
Although the calculation is straightforward, the intermediate
expressions are lengthy, and we omit them here. 
The intermediate
steps will be presented in~\cite{rf:KTY}.
Thus, for nonzero momentum, the trace-dependent part of the
ghost-number-three graviton vertex operator is BRST exact.
This argument does not apply at zero momentum, since the auxiliary
vector $\bar{k}^\mu$ cannot be introduced.

We next examine the ascendants related to $\omega_0^3$ through the
descent equations
\begin{align}
\bm{\delta}_\mathrm{B}{\omega_2}^1=-d{\omega_1}^2,
\qquad
\bm{\delta}_\mathrm{B}{\omega_1}^2=d{\omega_0}^3.
\end{align}
They are given by
\begin{align}
{\omega_2}^1
&=
-dz\wedge d\bar{z}\,
\frac{1}{4\pi}(\partial c-\bar{\partial}\tilde{c})
\varepsilon_{\mu\nu}
%\partial X^\mu\bar{\partial}X^\nu e^{ik\cdot X}
{\cal G}^{\mu\nu}
+\frac{1}{2}{\varepsilon^\mu}_{\mu}\,
\varDelta{\omega_2}^1(h),
\nn
{\omega_1}^2
&=
-\frac{1}{4\pi}(\partial c-\bar{\partial}\tilde{c})
\left\{
d\bar{z}\left(
c\varepsilon_{\mu\nu}\
%partial X^\mu \bar{\partial}X^\nu e^{ik\cdot X}
{\cal G}^{\mu\nu}
-\frac{\alpha'}{8}{\varepsilon^\mu}_{\mu}
\bar{\partial}^2\tilde{c}\,e^{ik\cdot X}
\right)
\right.
\nonumber\\
&\left.
-dz\left(
\tilde{c}\varepsilon_{\mu\nu}
%\partial X^\mu \bar{\partial}X^\nu e^{ik\cdot X}
{\cal G}^{\mu\nu}
-\frac{\alpha'}{8}{\varepsilon^\mu}_{\mu}
\partial^2c\,e^{ik\cdot X}
\right)
\right\}
+\frac{1}{2}{\varepsilon^\mu}_{\mu}\,
\varDelta{\omega_1}^2(h).
\end{align}
The additional terms $\varDelta{\omega_2}^1(h)$
and $\varDelta{\omega_1}^2(h)$ are given by
\begin{align}
\varDelta {\omega_2}^1(h)=&dz\wedge d\bar{z}
\left\{
\bar{\partial}^2 \tilde{c}C(h)-\partial^2c \tilde{C}(h)
\right\}
\label{eq:dW21h}
\\
\varDelta {\omega_1}^2(h)=&
d\bar{z}\left\{
c\partial^2 c \tilde{C}(h)
+\bar{\partial}^2 \tilde{c} c C(h)
+\frac{\alpha'}{2}
(\partial c+\bar{\partial}\tilde{c})\bar{\partial}^2\tilde{c}\,e(h)\right\}
\nn
&
\hspace{-1.3cm}
+dz\left\{
\tilde{c}\bar{\partial}^2 \tilde{c} C(h)
+\partial^2 c\tilde{c}  \tilde{C}(h)+\frac{\alpha'}{2}
(\partial c+\bar{\partial}\tilde{c})\partial^2c\,e(h)\right\},
\label{eq:dW12h}
\end{align}
where $C(h)$, $\tilde{C}(h)$, and $e(h)$ are defined by \eqref{eq:Ce_def}.

As in the zero-form case, we decompose the ascendants
into transverse-traceless and trace-dependent parts.

For the two-form, we write
\begin{align}
{\omega_2}^1
&=
\omega^1_{2{\rm T}}
+\frac{1}{2}{\varepsilon^\mu}_{\mu}\,
\omega^1_{2{\rm S}},
\\
\omega^1_{2{\rm T}}
&=-
\frac{1}{4\pi}(\partial c-\bar{\partial}\tilde{c})
dz\wedge d\bar{z}\,
\varepsilon'_{\mu\nu}\mathcal{G}^{\mu\nu},
\\
\omega^1_{2{\rm S}}
&=-
\frac{1}{4\pi}(\partial c-\bar{\partial}\tilde{c})
dz\wedge d\bar{z}\,
(k_\mu\bar{k}_\nu+k_\nu\bar{k}_\mu)
\mathcal{G}^{\mu\nu}
+\varDelta\omega_2^1(h).
\end{align}
Using the BRST transformations given above,
this trace-dependent part can be rewritten as
\begin{align}
\omega^1_{2{\rm S}}
=&\,\,
-{\bm\delta}_{\rm B}\Phi_2^0(F,\bar{k})
-d\,\Big[
\Phi_1^1(F,\bar{k})
\nonumber\\
&
\qquad
+\frac{i}{4\pi}(\partial c-\bar{\partial}\tilde{c})
\left(
dz\,\bar{k}\cdot A
-d\bar{z}\,\bar{k}\cdot\tilde{A}
\right)
\Big],
\label{eq:trace2form}
\end{align}
where the operators $\Phi_1^1(F,\bar{k})$ and
$\Phi_2^0(F,\bar{k})$ are defined by
\begin{align}
\Phi_1^1(F,\bar k)
=\,&
dz\left\{
\frac{4}{\alpha'}\tilde c\,
\bar{\varepsilon}^{(-)}_{\mu\nu}(F)\mathcal{G}^{\mu\nu}
-2i(\partial c+\bar{\partial}\tilde c)
\bar k_\mu F(k\cdot X)A^\mu
\right\}
\nonumber\\
&
\hspace{-1.2cm}
-d\bar z\left\{
\frac{4}{\alpha'}c\,
\bar{\varepsilon}^{(-)}_{\mu\nu}(F)\mathcal{G}^{\mu\nu}
+2i(\partial c+\bar{\partial}\tilde c)
\bar k_\mu F(k\cdot X)\tilde A^\mu
\right\},
\label{eq:Phi11_def}
\\
\Phi_2^0(F,\bar k)
=\,&
dz\wedge d\bar z\,
\frac{4}{\alpha'}
\bar{\varepsilon}^{(-)}_{\mu\nu}(F)\mathcal{G}^{\mu\nu}.
\label{eq:Phi20_def}
\end{align}
Thus, the trace-dependent part of the two-form operator
is BRST exact modulo a total derivative.

Similarly, for the one-form, we write
\begin{align}
{\omega_1}^2
&=
\omega^2_{1{\rm T}}
+\frac{1}{2}{\varepsilon^\mu}_{\mu}\,
\omega^2_{1{\rm S}},
\\
\omega^2_{1{\rm T}}
&=
-\frac{1}{4\pi}(\partial c-\bar{\partial}\tilde{c})
(d\bar{z}\,c-dz\,\tilde{c})\,
\varepsilon'_{\mu\nu}\mathcal{G}^{\mu\nu},
\\
\omega^2_{1{\rm S}}
&=
-\frac{1}{4\pi}(\partial c-\bar{\partial}\tilde{c})
\,\Bigg\{
(d\bar{z}\,c-dz\,\tilde{c})
(k_\mu\bar{k}_\nu+k_\nu\bar{k}_\mu)
\mathcal{G}^{\mu\nu}
\nonumber\\
&
-(d\bar{z}\,\bar{\partial}^2\tilde{c}
-dz\,\partial^2c)
\frac{\alpha'}{4}e^{ik\cdot X}
\Bigg\}
+\varDelta\omega_1^2(h).
\end{align}
Using the same BRST transformations, the trace-dependent
part becomes
\begin{align}
\omega^2_{1{\rm S}}
&=
{\bm\delta}_{\rm B}
\left[
\Phi_1^1(F,\bar{k})
+\frac{i}{4\pi}(\partial c-\bar{\partial}\tilde{c})
\left(
dz\,\bar{k}\cdot A
-d\bar{z}\,\bar{k}\cdot\tilde{A}
\right)
\right]
\nonumber\\
&\quad
+d\left[
\Phi_0^2(F,\bar{k})
+\frac{i}{4\pi}(\partial c-\bar{\partial}\tilde{c})
\left(
c\,\bar{k}\cdot A
-\tilde{c}\,\bar{k}\cdot\tilde{A}
\right)
\right].
\label{eq:trace1form}
\end{align}

Equations \eqref{eq:trace2form}, \eqref{eq:trace1form}, and
\eqref{eq:traceBRSTexact} show that, for nonzero momentum,
the
trace-dependent sector forms a trivial descent chain.
More precisely, the trace-dependent parts of the two- and
one-form operators are BRST exact modulo total derivatives, while the
zero-form operator is BRST exact itself. 
Thus,
the BRST triviality of the trace mode is compatible with
the entire descent structure.

As before, this conclusion relies on the existence of the auxiliary
null vector $\bar{k}^\mu$ satisfying $k\cdot\bar{k}=1$ and therefore
does not extend directly to zero momentum.

\section{Disk one-point amplitude for the graviton}

Having constructed the graviton vertex operator with ghost number
three, we now apply it to the disk one-point amplitude in the
presence of a D$p$-brane. 
We again impose the de Donder condition on the polarization tensor
$\varepsilon_{\mu\nu}$ without requiring tracelessness.

Following the dilaton tadpole calculation in~\cite{Kishimoto:2024yuw},
we regard the vertex operator as initially defined at the origin
of the unit disk through the state-operator correspondence
and map the unit disk to the upper half-plane.
Let $z'=f(z)$
denote the corresponding map as in~\cite{Kishimoto:2024yuw},
where $z$ is the coordinate on the
unit disk and $z'$ is that on the upper half-plane. 
We denote by $z_0=f(0)$ the image of the origin of the unit disk,
which is the insertion point of the vertex operator on the upper half-plane.
The disk
one-point amplitude is then obtained from the correlation function
\begin{align}
 {\cal A}_{\rm grav}
 =
 \frac{2}{\alpha'}\,
 i g_{\rm c} C_{D_2}
 \left\langle 0\right|
 f\circ{\omega_0}^3(0)
\left|0\right\rangle,
\end{align}
where $g_\mathrm{c}$ is the closed string coupling constant
and $C_{D_2}=1/(\alpha' g_\mathrm{o}^2)$, with $g_\mathrm{o}$
denoting the open string coupling constant.
The factor $2/\alpha'$ is included in the normalization of
the graviton vertex operator following the convention of
Polchinski~\cite{Polchinski:1998rq}.

Using the conformal transformation of ${\omega_0}^3$
and evaluating the ghost correlators as in the dilaton
case~\cite{Kishimoto:2024yuw}, we obtain, without yet specifying the
boundary conditions for the matter fields,
\begin{align}
 \left\langle 0\right|
 f\circ{\omega_0}^3(0)
\left|0\right\rangle
=\,&
-\frac{1}{2\pi}(z_0-\bar{z}_0)^2
 \left\langle 0\right|
\varepsilon_{\mu\nu}\partial X^\mu
\bar{\partial}X^\nu e^{ik\cdot X(z_0,\bar{z}_0)}
\left|0\right\rangle
\nonumber\\
\,&
+\frac{1}{2}{\varepsilon^\mu}_\mu\,\frac{\alpha'}{2\pi}\left\langle 0\right|
e^{ik\cdot X(z_0,\bar{z}_0)}
\left|0\right\rangle.
\end{align}

To verify that the ghost-number-three vertex operator reproduces
the expected amplitude, 
we consider a D$p$-brane
and impose the corresponding boundary conditions on the matter
fields.  
We decompose the spacetime directions $\mu=0,\cdots,25$ into
the directions $a=0,\cdots,p$ parallel to the D$p$-brane and
$i=p+1,\cdots,25$ transverse to it.
We take $X^a\ (a=0,\cdots,p)$ to satisfy Neumann
boundary conditions and $X^i\ (i=p+1,\cdots,25)$ to satisfy
Dirichlet boundary conditions.

Using these boundary conditions, the matter correlation functions
on the upper half-plane are evaluated as
\begin{align}
 -\frac{1}{2\pi}(z_0-\bar z_0)^2
&
 \left\langle 0\right|
 \varepsilon_{\mu\nu}\partial X^\mu
 \bar{\partial}X^\nu
 e^{ik\cdot X(z_0,\bar z_0)}
 \left|0\right\rangle
\nonumber\\
=\,\,&
(2\pi)^{p+1}\delta^{p+1}(k^a)\frac{\alpha'}{4\pi}\left(
{\varepsilon^a}_a-{\varepsilon^i}_i\right),
\end{align}
whereas
\begin{align}
\frac{1}{2} {\varepsilon^\mu}_\mu
\frac{\alpha'}{2\pi} 
&\left\langle 0\right|
 e^{ik\cdot X(z_0,\bar z_0)}
 \left|0\right\rangle
\nonumber\\
=\,\,&
(2\pi)^{p+1}\delta^{p+1}(k^a)\frac{\alpha'}{4\pi}\left(
{\varepsilon^a}_a+{\varepsilon^i}_i\right).
\end{align}
Substituting these expressions into the correlation function above,
the contributions proportional to ${\varepsilon^i}_i$ cancel
between the two terms.
Consequently, the disk one-point amplitude becomes
\begin{align}
 {\cal A}_{\rm grav}
 =
i(2\pi)^{p+1}\delta^{p+1}(k^a)\,
\frac{g_\mathrm{c}}{\pi \alpha' g_\mathrm{o}^2}\,{\varepsilon^a}_a,
\end{align}
which agrees, including the overall normalization, with
the graviton one-point amplitude obtained from the
D$p$-brane  effective action, using the normalization conventions
of~\cite{Polchinski:1998rq}.
The cancellation of the Dirichlet components also shows explicitly
that the trace-dependent term in ${\omega_0}^3$ is essential for
obtaining the correct amplitude.

This result illustrates a more general point: BRST triviality at
nonzero momentum does not by itself imply that the corresponding
sector can be discarded in a covariant description of zero-momentum
amplitudes.  It would be interesting to clarify further the relation
of this trace-dependent sector to the zero-momentum dilaton sector
and to its counterpart in closed string field theory.  These
questions will be addressed elsewhere.

%%%%%%%%%%%%%%%%%%%%%%%%%%%%%%%%%%%%%%%
\section*{Acknowledgments}
We would like to thank S. Seki for useful discussions.
T.~T.~was supported in part by JSPS KAKENHI Grant Numbers
\#23K03388 and \#26K22332.  M.~Y.~was supported by the SanDisk
Scholarship and the Initiative for Realizing Diversity in the Research Environment.
%%%%%%%%%%%%%%%%%%%%%%%%%%%%%%%%%%%%%%%%%%%%%%%%%%%%%%

%\section*{Declaration of generative AI and AI-assisted technologies
%in the manuscript preparation process}
%
%During the preparation of this work the authors used ChatGPT in order
%to improve the language and readability of the manuscript.
%After using this tool/service, the authors reviewed and edited the
%content as needed and take full responsibility for the content of
%the published article.

%%%%%%%%%%%%%%%%%%%%%%%%%%%%%%%%%%%%%%%%%%%%%%%%%%%%%

 \end{document}